# Scalable and efficient grating couplers on low-index photonic platforms enabled by cryogenic deep silicon etching

Emma Lomonte[1,2,3†], Maik Stappers[1,2,3†], Linus Krämer[1,2,3,4], Wolfram H.P. Pernice[1,2,3,4*], and Francesco Lenzini[1,2,3*]

[1] *Institute of Physics, University of Münster, Wilhelm-Klemm-Straße 10, 48149 Münster, Germany*
[2] *CeNTech – Center for Nanotechnology, Heisenbergstraße 11, 48149 Münster, Germany*
[3] *SoN – Center for Soft Nanoscience, Busso-Peus-Straße 10, 48149 Münster, Germany*
[4] *Heidelberg University, Im Neuenheimer Feld 227, 69120 Heidelberg, Germany*

* *wolfram.pernice@uni-muenster.de*, *lenzini@uni-muenster.de*

[†]*Equally contributing authors.*

Efficient fiber-to-chip couplers for multi-port access to photonic integrated circuits are paramount for a broad class of applications, ranging, e.g., from telecommunication to photonic computing and quantum technologies. While grating-based approaches are convenient for out-of-plane access and often desirable from a packaging point of view, on low-index photonic platforms, such as silicon nitride or thin-film lithium niobate, the limited grating strength has thus far hindered the achievement of coupling efficiencies comparable to the ones attainable in silicon photonics. Here we present a flexible strategy for the realization of highly efficient grating couplers on low-index photonic platforms. To simultaneously reach a high scattering efficiency and a near-unitary modal overlap with optical fibers, we make use of self-imaging gratings designed with a negative diffraction angle. To ensure high directionality of the diffracted light, we take advantage of a metal back-reflector patterned underneath the grating structure by cryogenic deep reactive ion etching of the silicon handle. Using silicon nitride as a testbed material, we experimentally demonstrate coupling efficiency up to -0.55 dB in the telecom C-band with near unity chip-scale device yield.

**INTRODUCTION**

In the last decades, photonic integrated circuits (PICs) have captured a growing interest in several areas spanning, e.g., from optical communication [1,2] to quantum technologies [3,4] and artificial intelligence [5–7] thanks to their potential for miniaturization and intrinsically large bandwidth. By leveraging ever-more mature fabrication processes, scale and complexity of PICs are steadily increasing [8]. To connect photonic chips designated to deliver different tasks in a hybrid architecture or to minimize the power requirements on active off-chip components such as lasers, the availability of highly efficient fiber-to-chip coupling interfaces is a crucial requisite [9]. Fiber-to-chip interconnects with a coupling efficiency approaching unity are especially critical components for applications in photonic quantum computing, where, regardless of the computation being based on discrete-variable (DV) or continuous-variable (CV) encoding, any source of loss can either severely limit the scalability of the system or introduce unacceptable levels of noise in the quantum states [10–12].

Since the rise of silicon photonics, the size mismatch between optical fibers characterized by an approximately 8 μm large core (at λ=1550 nm) and waveguides with sub-micron cross sections has made the realization of efficient coupling interfaces particularly challenging. At present, the two most employed strategies to couple light into and out of PICs are either edge couplers - where light must be guided to the facet of the chip and then in-plane into the optical fiber -, or grating couplers, which can diffract light out-of-plane and therefore couple to fiber array units placed atop the chip [9]. Edge couplers are currently the predominant choice for applications requiring broad bandwidth and polarization insensitivity. Conversely, grating couplers typically display a narrower bandwidth and are sensible to the polarization of the input light. However, they naturally offer more flexible positioning on-chip, enabling a denser integration of components and large-scale prototyping. Moreover, while edge couplers often require the use of lensed or ultra-high numerical aperture (UHNA) fibers, grating couplers can more easily achieve high coupling efficiency with standard single-mode (SM) optical fibers [13]. Out-of-plane couplers are also the favourite choice for photonic packaging thanks to the availability of a larger surface area for fiber bonding [14].

Conventional grating couplers typically offer limited coupling efficiency, mainly capped by the imperfect directionality of the diffraction process. For example, assuming a perfectly symmetric material stack where a fully etched grating is immersed in infinite upper and bottom lower refractive index claddings, the theoretical maximum coupling efficiency is -3 dB since upward and downward diffraction are completely equivalent. Extensive efforts have been devoted to enhancing the grating directionality, i.e., the fraction of light radiated in the upwards direction compared to the total one diffracted by the grating, in the past years. Current solutions employ e.g., metal mirrors or distributed Bragg reflectors (DBRs) [15–19], dual-etch gratings [20–22], or bi-level grating structures [22–24]. Grating couplers with a metal back-reflector implemented on silicon-on-insulator (SOI) have been proven so far as the most effective choice, unlocking coupling efficiencies up to ≃-0.5 dB at telecom wavelength [15,16,18].

Although SOI is still - and will likely remain - the favorite choice for the optical communication sector, its limited transparency range, presence of two-photon absorption, and absence of second-order nonlinearities have motivated the development of PICs also on different materials. In this context, photonic platforms characterized by a refractive index ≃2 at telecom wavelength, such as silicon nitride (SiN) or thin-film lithium niobate (TFLN), have stood out as new leading alternatives for applications involving the generation and manipulation of quantum states of light [23–28], microcomb generation [29–31], and high-speed optical signal processing [32–34]. Despite the refractive index substantially smaller than the one of silicon (n≃3.5 at λ=1550 nm), their index contrast is still sufficient for realizing compact PICs with bending radii of only few tens of micron. However, the lower refractive index comes at the price of a reduced grating strength, making the realization of highly efficient grating couplers particularly challenging. Indeed, for this class of photonic platforms, the best gratings demonstrated so far achieve a coupling efficiency up to -0.89 dB for the case of TFLN [35], and up to -1.17 dB for the case of SiN [17]. Such values are substantially lower than the ones obtained at telecom wavelength on SOI [15,16,18].

Here we propose and experimentally demonstrate a flexible strategy for the realization of highly efficient grating couplers on low-index platforms. To overcome the limitation of reduced grating strength, we make use of apodized grating couplers designed to display a negative diffraction angle instead of a more conventional positive one [22,36–38]. This allows us to obtain a high scattering efficiency and diffract all the incoming light, and, simultaneously, to achieve a focusing effect of the Gaussian-like diffracted beam at a distance of ≃100-200 µm away from the circuit plane to precisely match the spatial mode supported by an optical fiber [38]. To obtain high directionality of the diffraction process, we realize a metal back-reflector underneath the buried oxide layer by cryogenic deep reactive ion etching of the silicon handle. For a proof-of-concept, we choose SiN as a testbed material, that has attracted growing attention for applications involving both classical and quantum states of light because of its wide transparency range, ultra-low propagation loss, and absence of two-photon absorption [39]. With fully etched grating couplers patterned on a 330 nm thick SiN film, we achieve a coupling efficiency up to -0.55 dB at telecom wavelength, a value comparable with state-of-the art gratings implemented in silicon photonics [15,16,18]. Importantly, we stress that the proposed strategy relies neither on an optimal material refractive index, nor on optimally engineered film thickness and etching depth. Thus, our approach can be translated to many other low-index platforms, such as TFLN, aluminum nitride, or tantalum pentoxide [40–42].

**THEORY AND SIMULATIONS**

The coupling efficiency (CE) of a grating coupler can be modeled as the product of three different contributions:

$$CE=\eta_1 \cdot \eta_2 \cdot \eta_3,$$

where $\eta_1$ is the scattering efficiency of the grating structure, $\eta_2$ is the bidimensional overlap integral between the diffracted optical beam and the fundamental light mode supported by the optical fiber, and $\eta_3$ is the directionality of the diffraction process. The scattering efficiency $\eta_1$ and the overlap integral $\eta_2$ can be boosted to values close to 100% via an accurate design of the surface grating teeth. In contrast, high directionality $\eta_3$ requires proper engineering of the material stack for reflecting upward the light that otherwise would leak into the substrate.

Elaborating further, near-unity scattering efficiencies can be straightforwardly attained in all photonic platforms by designing grating couplers with a large enough number of grating teeth. In the case of low-index platforms the limited grating strength can be bypassed simply by using longer gratings. Achieving an optimal 2D overlap integral requires fulfilling two conditions: *i)* the diffracted beam must feature a Gaussian-like profile as does the fundamental mode of optical fibers; *ii)* the mode field diameters (MFDs) of the radiated beam and the fundamental fiber mode must be identical along both the longitudinal and the transverse directions. While the former requirement can be met by exploiting apodization of the filling factor such that the highest amount of optical power is diffracted at the center of the grating rather than at its beginning [43], the attainment of the latter one can easily become challenging in grating couplers characterized by a moderate grating strength, whereas ≥40 µm long structures are needed to diffract most of the propagating light [22,38], thus greatly deteriorating the overlap between the two optical modes involved. As we shall see below, this conflict can be overcome using self-imaging grating couplers designed with a negative diffraction angle, where, by taking advantage of the focusing effect of the upward diffracted light, high values for both $\eta_1$ and $\eta_2$ can be simultaneously achieved [22,36–38].

To elucidate this approach, we design air-clad fully etched grating couplers patterned into a 330 nm thick silicon nitride film on a 3.33 µm thick buried oxide layer (see Fig. 1a,b). To enhance the directionality of the grating, an aluminum (Al) film is placed underneath the buried oxide layer acting as a metal back-reflector. The grating couplers are designed for operation in the telecom C-band with TE-polarized light, while a negative diffraction angle is chosen to achieve a focusing effect for the upward diffracted light. To numerically optimize the grating coupler along its longitudinal direction, 2D finite-domain time-difference (FDTD) simulations were performed with MEEP, an open-source software package for electromagnetics simulations [44], while analytical formulas for Gaussian beam propagation were used to calculate the remaining grating parameters along the transverse direction. For both numerical optimization and analytical design, we follow the strategy developed in our previous work and described in detail in Ref. [38].

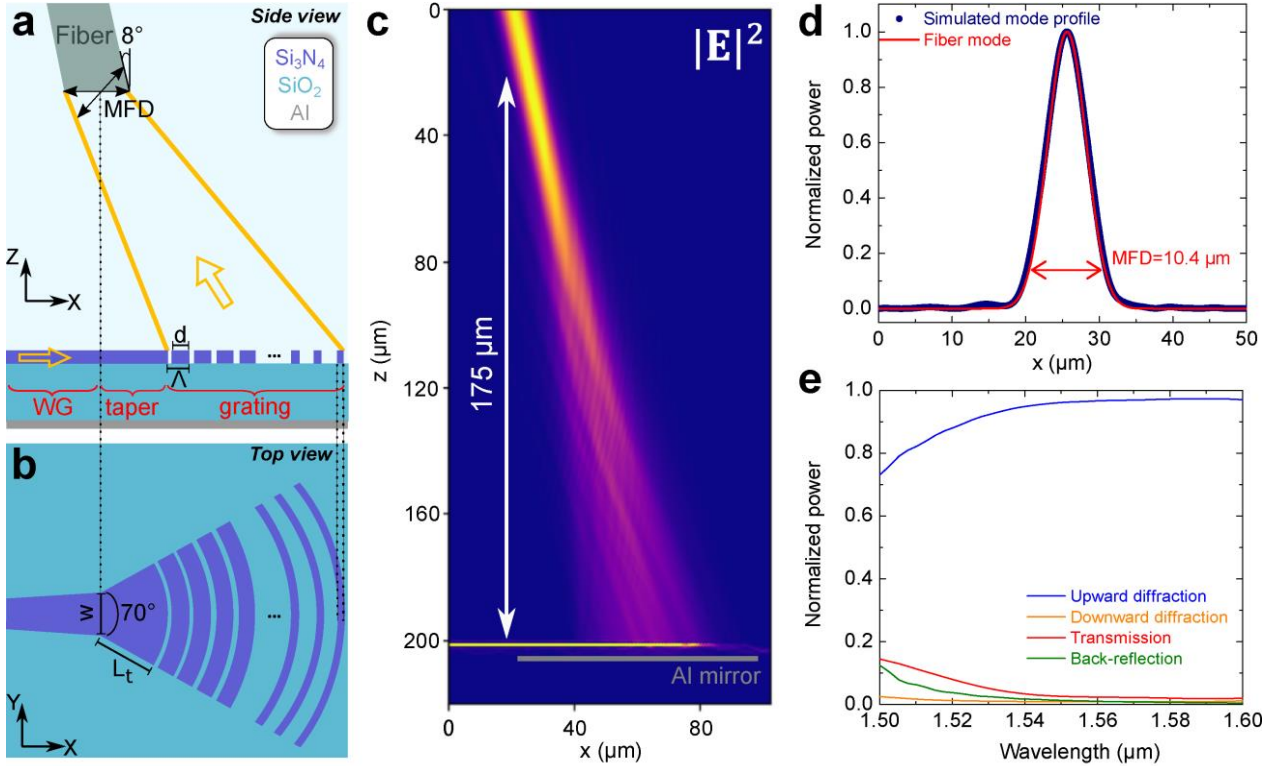

**Figure 1. (a)** Schematic illustration of the grating coupler along its longitudinal direction. The grating coupler features a grating period Λ and a linear apodization on the filling factor FF=d/Λ, such that the light propagating through the structure (yellow arrow) is partly radiated upward with a Gaussian-like shape and focused at a given distance from the circuitry plane. An optical fiber with the MFD of its fundamental mode is also drawn in correspondence of the beam waist of the radiated beam. Highlighted with red labels are also the several portions of the device: the waveguide (WG), the grating and the taper connecting these two elements. **(b)** Top view of the self-imaging apodized grating coupler. The taper features an opening angle of 70° and a length $L_t$ equal to the distance between the first grating tooth and the longitudinal coordinate of the beam waist. At the intersection with the taper, the waveguide is characterized by a properly engineered width w that allows to match the desired MFD also in the transverse direction, in correspondence of the focal point of the diffracted beam. **(c)** 2D FDTD simulation of a grating coupler in the outcoupling scheme at λ=1550 nm. The colormap in the figure (a.u.) is the computed optical power $|E|^2$ of the diffracted beam. The radiated beam features a diffraction angle equal to -12° and a focusing effect at a height of ≃175 μm from the circuit plane. **(d)** Simulated mode field profile in correspondence of the beam waist (blue curve). The red trace is a Gaussian function with MFD=10.4 μm (MFD of SMF-28 fibers at λ=1550 nm). **(e)** Poynting flux spectra computed in the 1500-1600 nm wavelength range for upward-(blue) and downward-diffracted (orange) light, for light transmitted at the end of the grating structure (red), and for the back-reflected light (green). The optical power is normalized to the one at the input of the grating.

A side view of the grating coupler is shown in Fig. 1a. The grating is characterized by a constant grating period Λ, and a linear apodization of the filling factor FF=d/Λ. The apodization of the filling factor has two distinct functions (see Ref. [38] for more details): on the one hand, it ensures that the diffracted beam has a Gaussian-like profile; on the other hand, as the diffraction angle depends on the local filling factor (with a smaller filling factor resulting in a larger negative diffraction angle) it also enables to achieve a focusing effect along the longitudinal direction of the upward diffracted beam. Compatible with the resolution of our fabrication process, the initial filling factor is kept as large as possible to minimize back-reflections into the waveguide due to the index mismatch between the input waveguide and the grating. The number of grating teeth is instead kept constant to ensure a scattering efficiency close to 100%. The grating period and the final filling factor are then optimized with 2D FDTD simulations such that the upward radiated beam is characterized by a diffraction angle of -12° - to match, after taking into account Fresnel refraction, the 8° polishing angle of our fiber array - as well as a mode field diameter

(MFD) in correspondence of the focal point as close as possible to the targeted value of 10.4 µm (MFD of SMF-28 fibers at λ=1550 nm).

In Fig. 1b we show a top view of the grating coupler. The grating teeth are shaped as a series of concentric circular arcs featuring a radius equal to the distance from the input waveguide. Taking advantage of the negative diffraction angle, the circular shape enables to achieve a focusing effect also along the transverse direction of the diffracted beam [22,36–38]. To connect the waveguide to the circular arcs, a short transition taper is employed. Its wide opening angle of 70° ensures that the light mode is no longer confined along the transverse direction of the grating and freely propagates in the transition region, while its length $L_t$ is chosen equal to the distance between the X-coordinate of the beam waist of the upward diffracted beam and the first grating tooth. This design choice guarantees that the longitudinal and the transverse focal points of the upward diffracted beam occur at the same position [37,38]. The waveguide width w at the intersection with the taper is chosen to match the targeted MFD also in the transverse direction of the propagating beam and can be estimated using standard analytical formulas for Gaussian beam propagation in free-space and dielectric media (see Ref. [38] for more details).

Fig. 1c displays the optical power of the upward diffracted beam, computed with 2D FDTD simulations for the optimized grating couplers parameters. The simulation is performed in the outcoupling regime, by injecting a continuous wave light source with 1550 nm wavelength at the input of the grating. The grating consists of 50 apodized grating teeth, followed by 10 uniform ones which are introduced to minimize possible back-reflections at the end of the structure. The grating period Λ is equal to 890 nm, and an initial and final filling factor of 95% and 60% are used, respectively. Fig. 1d shows the computed optical power within the focal point, located at a height of $\simeq$175 µm from the circuit plane. A direct comparison with a Gaussian beam with 10.4 µm MFD (see the red curve in the figure) leads to a 1D overlap with the mode supported by SMF-28 fibers equal to $\simeq$98%. For this set of grating parameters, an optimal taper length $L_t \simeq$ 15 µm, and a waveguide width w $\simeq$ 2 µm at the intersection with the taper are estimated.

In Fig. 1e we show the Poynting flux spectra computed in the 1500-1600 nm wavelength range for the upward and downward diffracted light (blue curve and orange curve, respectively), the light transmitted at the end of the grating (red curve), and the light back-reflected into the waveguide (green curve). At wavelengths larger than 1.55 µm, a directivity >97% is achieved. We also note that, as expected, only a negligible amount of light is transmitted through the end of the grating, while back-reflections into the waveguide are suppressed by more than 20 dB. Assuming a perfect modal overlap in the transverse direction of the propagating beam, for our gratings we estimate a peak coupling efficiency equal to $\simeq$-0.2 dB. We point out that a large fraction of this value is capped by the limited reflectivity of the aluminum mirror ($\simeq$96% at telecom wavelength), and even higher coupling efficiencies could be possible with a non-metallic back-reflector displaying a higher reflectivity.

**FABRICATION**

For the experimental implementation of the proposed grating couplers, the fabrication workflow is divided into two main parts: at first, we realize the metal mirror by processing the backside of the chip and, afterwards, we pattern the photonic circuitry. Our starting point is a 20x20 mm$^2$ die, consisting of a 330 nm thick SiN film deposited via low-pressure chemical vapor deposition (LPCVD) on a 3.33 µm thick $SiO_2$ layer, thermally grown on a 525 µm thick Si handle. Prior to nanoprocessing, the die is annealed at 1100° C for 4 hours in nitrogen atmosphere to reduce the absorption loss of the nitride film [45]. Subsequently, we polish the Si handle down to a thickness of approximately 200 µm, to enable good coverage of the fabricated membranes with the metal layer during the later deposition process.

A dense matrix of $SiO_2$ membrane windows characterized by a width of approximately 150 µm (see Fig.2a) are patterned on the backside of the sample by photolithography with a 30 µm thick SU-8 photoresist, followed by cryogenic deep reactive ion etching (DRIE) of the silicon handle. We perform DRIE in a $SF_6+O_2$ gas mixture using an ICP-RIE etching tool (Oxford Instrument PlasmaPro 100), whose table temperature is cooled down to -100° C with liquid nitrogen [46,47]. The DRIE process is preferred over more common backside wet etching methods [15,18] because of the possibility of realizing membrane windows with almost perfectly vertical sidewalls. Indeed, when performing wet etching with KOH or TMAH solutions, the etched Si sidewalls are characterized by an angle of approximately 54°, greatly limiting the maximum achievable size of the membrane. For example, membranes only 50 µm wide results in openings as large as 1 mm at the exposed surface of the Si substrate [18], not only largely affecting the stability of the photonic die and limiting the density of PICs, but also making it impossible to accommodate grating couplers with a sizable number of grating teeth, as needed when working with low-index photonic platforms. To prevent micromasking effects and obtain a smooth $SiO_2$ surface, the membranes are not fully etched by DRIE and the last $\simeq$10 µm of silicon are removed by wet etching with TMAH solution.

After DRIE, the PICs are fabricated in the SiN front side by electron-beam lithography (EBL) with a negative resist (ArN-7520), followed by reactive ion etching with a $CHF_3+O_2$ chemistry. We use the scanning electron microscope of our EBL system for precisely aligning the grating couplers relative to the fabricated membrane windows. A photolithography system with back-side alignment would enable to reverse the process and realize the membrane windows after waveguide patterning, increasing the robustness of our approach to workflows requiring multiple fabrication steps. Lastly, a ≃150 nm thick aluminum film is deposited on the backside of the sample by DC magnetron sputtering. At the end of the fabrication process, we observed a high chip-scale device yield of ≃97%. In Fig. 2b we show a microscope image of a completed sample with several grating couplers patterned atop the metal mirrors. The devices shown in the picture consist of two identical grating couplers separated by a distance of 254 µm (equal to twice the pitch of our in-house fiber array). The gratings are connected by a waveguide with a width of 1.3 µm to ensure single-mode operation in the 1550 nm wavelength range. A 50 µm long adiabatic taper connects the single-mode waveguide to the larger waveguide at the input of the grating as described above. Lithographic tuning of the various grating parameters is performed on chip, using the ones optimized with numerical simulations as a starting guess.

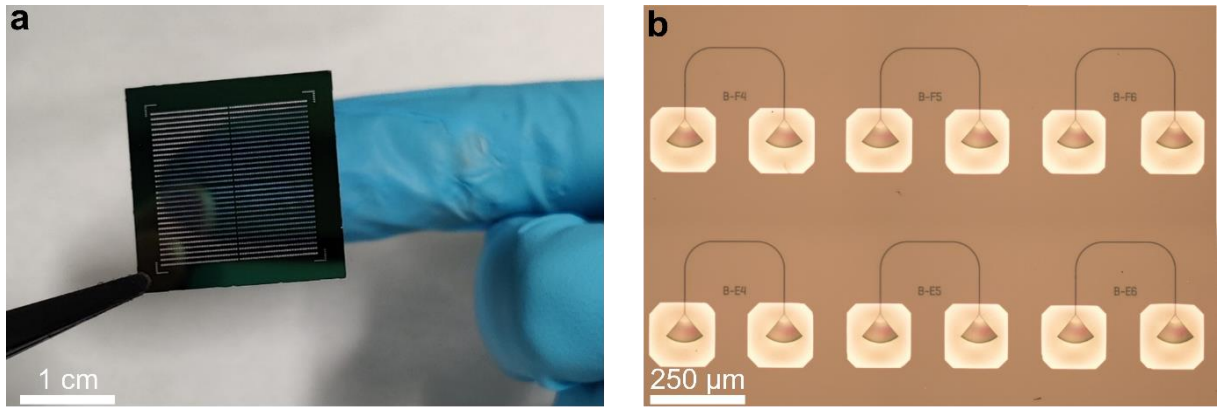

**Figure 2. (a)** Photograph of a 20x20 mm² die, where a dense matrix of membrane windows is fabricated via back-side etching of the silicon handle. The fabricated membranes contain only a few micrometers of material and are therefore transparent. **(b)** Optical microscope image of a completed sample, showing several grating couplers patterned atop the metal mirrors.

**RESULTS**

To determine the efficiency of the fabricated couplers, we use a V-groove fiber array equipped with SMF-28 fibers and featuring an 8° polishing angle. A 3-axis linear stage is used for optimizing the coupling efficiency in the X-Y-Z directions, while a goniometric and a rotation stage are used for fine-tuning the inclination angle of the array with respect to the surface of the photonic die as well as the angle of the chip in the X-Y plane, respectively. Continuous-wave laser light from an external tunable laser source (TSL-550 Santec laser with 1500-1620 nm tuning range) is guided through a 3-paddle polarization controller to ensure that TE-polarized light is coupled into the PIC via one of the two accessible grating couplers. The transmission spectra are acquired by sweeping the wavelength of the laser source and by recording the light exiting from the other coupler with a photoreceiver.

The efficiency per coupler is estimated as the square root of the transmission of a device consisting of two identical grating couplers connected by a short waveguide (see Fig. 2b). To precisely evaluate the coupling efficiency of the gratings, the insertion loss of the two channels of our fiber array - which includes fiber connector loss (≃0.3 dB) and Fresnel loss at the output facet of the array (≃0.15 dB) - are preliminary characterized with an optical power meter and calibrated out from the transmission measurement. We note that Fresnel loss can be straightforwardly eliminated with an anti-reflection coating - a service commercially available from several fiber array manufacturers -, while a complete suppression of the connector loss would unavoidably require the use of fiber splicing. Propagation loss in the waveguide connecting the two couplers, estimated equal to ≃0.6 dB/cm by measuring the Q-factor of ring resonators fabricated on the same chip, are assumed to be negligible for the calculation of the coupling efficiency.

In Fig. 3.3a we report the coupling efficiency measured for the best device fabricated on chip as a function of the input laser wavelength. The measurement is performed by optimizing the transmission with a 1565 nm alignment wavelength, and by sweeping the laser frequency at a fixed fiber position. By fitting the data with a Gaussian function (see the inset in the figure) a peak coupling efficiency of -0.55 dB, corresponding to 88% in a linear scale, is determined. The ripples visible in the inset are likely due to small back-reflections of the fabricated gratings, generating a cavity effect in the short waveguide connecting the two couplers. Their small visibility (≃0.05 dB) indicates high suppression of back-reflections into the waveguides, as correctly predicted from numerical simulations (see Fig. 1e).

A drawback of self-imaging grating couplers is that light beams with different central wavelength are diffracted at slightly different angles and thus display a non-negligible spatial spreading in the focal point. This decreases the bandwidth measured at a fixed fiber position (3 dB bandwidth $\simeq$ 50 nm for the device of Fig. 3a) compared with more conventional grating configurations, in which the fiber is located at a close distance to the surface of the photonic die. However, as shown in Fig. 3b, a high coupling efficiency can be recovered in the full telecom C-band by tuning the fiber position and optimizing the transmission with a different alignment wavelength. In Fig. 3c we show, for the same device, the peak coupling efficiencies measured in the 1500-1600 nm wavelength range by tuning the alignment wavelength at steps of 5 nm. In good agreement with the flux spectra of Fig. 1e, we observe a fairly flat coupling efficiency in the 1550-1580 nm wavelength range, which gradually decreases as the alignment wavelength is tuned toward smaller and larger values. From these data, an effective 1 dB bandwidth of $\simeq$ 100 nm can be estimated for the tested gratings.

Finally, to study the reproducibility of our results, in Fig. 3d we also report the peak coupling efficiencies measured for the 10 identical devices that were fabricated on the same chip, obtained with the same procedure described for recording the data of Fig. 3a. Notably, except for a single device, all the others consistently achieve a coupling efficiency >-1 dB, with an average value <CE>=83% ± 3% (-0.81 dB ± 0.16 dB).

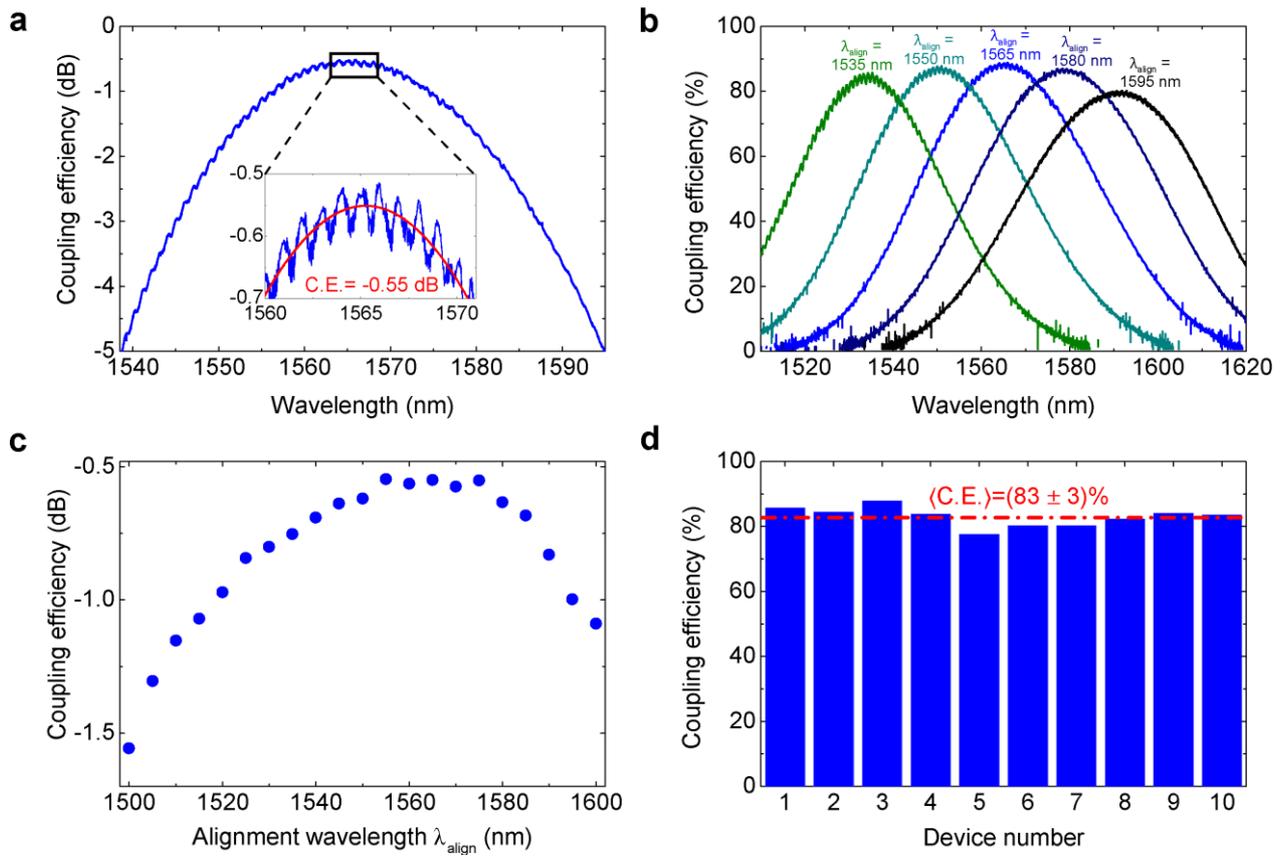

**Figure 3. (a)** Coupling efficiency of the best device fabricated on chip, plotted on a logarithmic scale. The transmission is optimized with a 1565 nm alignment wavelength. The laser is swept at a fixed fiber position. The inset shows the same curve, with a close-up around the maximum. The red curve is a Gaussian function used to fit the experimental data. A maximum coupling efficiency of -0.55 dB has been extracted. **(b)** For the same device, the coupling efficiency spectrum is reported for five different alignment wavelengths in the range of 1535-1595 nm. A peak coupling efficiency larger than 80% is determined for all cases. **(c)** For the same device, the peak coupling efficiency of the grating coupler is reported as a function of the alignment wavelength. The alignment wavelength is varied in steps of 5 nm in the range of 1500-1600 nm. **(d)** Peak coupling efficiencies measured for ten identical devices patterned on the same photonic die. An average coupling efficiency equal to (83 ± 3)% is determined. Device n.3 is the one used for the measurements of Fig. 3a-c.

## DISCUSSION

We have presented a flexible approach for the realization of highly efficient grating couplers, which, in principle, can be applied to any low-index photonic platform independently of the chosen film thickness and grating etching depth. With fully etched grating couplers fabricated on a 330 nm thick silicon nitride film, we have experimentally achieved a coupling efficiency up to -0.55 dB in the telecom C-band. We highlight that this value, measured on a low-index platform, not only is comparable with the best ones obtained at telecom wavelength for grating couplers implemented on silicon-on-insulator [15,16,18], but it is also competitive with state-of-the-art edge couplers making use of lensed or UHNA fibers [13].

It is currently an open question, and it will be a subject of future studies, how to further improve the coupling efficiency of our gratings toward the theoretical limit predicted by numerical simulations. For example, the relatively large variation in the coupling efficiency of identical devices suggests an improvable quality of the fabricated metal mirrors, whose deposition process was not extensively optimized in this work. Another possible factor limiting the efficiency of our couplers is related to the fact that commercially available V-groove fiber arrays typically display an average fiber core displacement in the range of 0.5-1 µm. In our case, this problem was minimized by measuring the coupling efficiency for different pairs of fibers and picking up the two displaying the best result. In future implementations, this issue could be overcome using two independently adjustable fibers for input and output coupling. Alternatively, if the position of the fiber cores can be precisely measured in advance, the separation between the grating couplers could be readjusted accordingly.

Overall, we believe that our approach opens a practical path toward the realization of vertical couplers featuring coupling efficiency approaching unity, as critically required for the implementation of scalable photonic technologies.

## ACKNOWLEDGEMENTS

We acknowledge support by the Deutsche Forschungsgemeinschaft (DFG) through the Münster Nanofabrication Facility (MNF) and CRC 1459. We also acknowledged funding by the European Union (Grant agreement IDs 101017237, 101046878 and 101080173) and the BMBF (PhoQuant, Hyphone, HybridQToke).

## REFERENCES

[1] W. Shi, Y. Tian, and A. Gervais, *Scaling Capacity of Fiber-Optic Transmission Systems via Silicon Photonics*, Nanophotonics **9**, 4629 (2020).

[2] S. Bernabé, Q. Wilmart, K. Hasharoni, K. Hassan, Y. Thonnart, P. Tissier, Y. Désières, S. Olivier, T. Tekin, and B. Szelag, *Silicon Photonics for Terabit/s Communication in Data Centers and Exascale Computers*, Solid State Electron **179**, 107928 (2021).

[3] J. Wang, F. Sciarrino, A. Laing, and M. G. Thompson, *Integrated Photonic Quantum Technologies*, Nat Photonics **14**, 273 (2020).

[4] G. Moody et al., *2022 Roadmap on Integrated Quantum Photonics*, Journal of Physics: Photonics **4**, 012501 (2022).

[5] Y. Shen et al., *Deep Learning with Coherent Nanophotonic Circuits*, Nat Photonics **11**, 441 (2017).

[6] J. Feldmann, N. Youngblood, C. D. Wright, H. Bhaskaran, and W. H. P. Pernice, *All-Optical Spiking Neurosynaptic Networks with Self-Learning Capabilities*, Nature **569**, 208 (2019).

[7] B. J. Shastri, A. N. Tait, T. Ferreira de Lima, W. H. P. Pernice, H. Bhaskaran, C. D. Wright, and P. R. Prucnal, *Photonics for Artificial Intelligence and Neuromorphic Computing*, Nat Photonics **15**, 102 (2021).

[8] D. Thomson et al., *Roadmap on Silicon Photonics*, Journal of Optics **18**, 073003 (2016).

[9] R. Marchetti, C. Lacava, L. Carroll, K. Gradkowski, and P. Minzioni, *Coupling Strategies for Silicon Photonics Integrated Chips [Invited]*, Photonics Res **7**, 201 (2019).

[10] M. V Larsen, C. Chamberland, K. Noh, J. S. Neergaard-Nielsen, and U. L. Andersen, *Fault-Tolerant Continuous-Variable Measurement-Based Quantum Computation Architecture*, PRX Quantum **2**, 30325 (2021).

[11] M. Varnava, D. E. Browne, and T. Rudolph, *How Good Must Single Photon Sources and Detectors Be for Efficient Linear Optical Quantum Computation?*, Phys Rev Lett **100**, 60502 (2008).

[12] E. J. Bourassa et al., *Blueprint for a Scalable Photonic Fault-Tolerant Quantum Computer*, Quantum **5**, 392 (2021).

[13] X. Mu, S. Wu, L. Cheng, and H. Y. Fu, *Edge Couplers in Silicon Photonic Integrated Circuits: A Review*, Applied Sciences.

[14] T. Tekin, *Review of Packaging of Optoelectronic, Photonic, and MEMS Components*, IEEE Journal of Selected Topics in Quantum Electronics **17**, 704 (2011).

[15] W. S. Zaoui, A. Kunze, W. Vogel, M. Berroth, J. Butschke, F. Letzkus, and J. Burghartz, *Bridging the Gap between Optical Fibers and Silicon Photonic Integrated Circuits*, Opt Express **22**, 1277 (2014).

[16] Y. Ding, C. Peucheret, H. Ou, and K. Yvind, *Fully Etched Apodized Grating Coupler on the SOI Platform with -0.58 DB Coupling Efficiency*, Opt Lett **39**, 5348 (2014).

[17] S. Nambiar, P. Ranganath, R. Kallega, and S. K. Selvaraja, *High Efficiency DBR Assisted Grating Chirp Generators for Silicon Nitride Fiber-Chip Coupling*, Sci Rep **9**, 18821 (2019).

[18] N. Hoppe, W. S. Zaoui, L. Rathgeber, Y. Wang, R. H. Klenk, W. Vogel, M. Kaschel, S. L. Portalupi, J. Burghartz, and M. Berroth, *Ultra-Efficient Silicon-on-Insulator Grating Couplers With Backside Metal Mirrors*, IEEE Journal of Selected Topics in Quantum Electronics **26**, 1 (2020).

[19] H. Zhang, C. Li, X. Tu, J. Song, H. Zhou, X. Luo, Y. Huang, M. Yu, and G. Q. Lo, *Efficient Silicon Nitride Grating Coupler with Distributed Bragg Reflectors*, Opt Express **22**, 21800 (2014).

[20] D. Benedikovic et al., *High-Directionality Fiber-Chip Grating Coupler with Interleaved Trenches and Subwavelength Index-Matching Structure*, Opt Lett **40**, 4190 (2015).
[21] D. Benedikovic et al., *L-Shaped Fiber-Chip Grating Couplers with High Directionality and Low Reflectivity Fabricated with Deep-UV Lithography*, Opt Lett **42**, 3439 (2017).
[22] Y. Chen, T. Domínguez Bucio, A. Z. Khokhar, M. Banakar, K. Grabska, F. Y. Gardes, R. Halir, Í. Molina-Fernández, P. Cheben, and J.-J. He, *Experimental Demonstration of an Apodized-Imaging Chip-Fiber Grating Coupler for Si3N4 Waveguides*, Opt Lett **42**, 3566 (2017).
[23] Y. Zhang, M. Menotti, K. Tan, V. D. Vaidya, D. H. Mahler, L. G. Helt, L. Zatti, M. Liscidini, B. Morrison, and Z. Vernon, *Squeezed Light from a Nanophotonic Molecule*, Nat Commun **12**, 2233 (2021).
[24] J. M. Arrazola et al., *Quantum Circuits with Many Photons on a Programmable Nanophotonic Chip*, Nature **591**, 54 (2021).
[25] C. Taballione, M. C. Anguita, M. de Goede, P. Venderbosch, B. Kassenberg, H. Snijders, D. Smith, J. P. Epping, R. van der Meer, and P. W. H. Pinkse, *20-Mode Universal Quantum Photonic Processor*, ArXiv Preprint ArXiv:2203.01801 (2022).
[26] R. Nehra, R. Sekine, L. Ledezma, Q. Guo, R. M. Gray, A. Roy, and A. Marandi, *Few-Cycle Vacuum Squeezing in Nanophotonics*, Science (1979) **377**, 1333 (2022).
[27] E. Lomonte, M. A. Wolff, F. Beutel, S. Ferrari, C. Schuck, W. H. P. Pernice, and F. Lenzini, *Single-Photon Detection and Cryogenic Reconfigurability in Lithium Niobate Nanophotonic Circuits*, Nat Commun **12**, 6847 (2021).
[28] P. I. Sund, E. Lomonte, S. Paesani, Y. Wang, J. Carolan, N. Bart, A. D. Wieck, A. Ludwig, L. Midolo, and W. H. P. Pernice, *High-Speed Thin-Film Lithium Niobate Quantum Processor Driven by a Solid-State Quantum Emitter*, ArXiv Preprint ArXiv:2211.05703 (2022).
[29] T. Herr, V. Brasch, J. D. Jost, C. Y. Wang, N. M. Kondratiev, M. L. Gorodetsky, and T. J. Kippenberg, *Temporal Solitons in Optical Microresonators*, Nat Photonics **8**, 145 (2014).
[30] B. Stern, X. Ji, Y. Okawachi, A. L. Gaeta, and M. Lipson, *Battery-Operated Integrated Frequency Comb Generator*, Nature **562**, 401 (2018).
[31] A. S. Raja et al., *Electrically Pumped Photonic Integrated Soliton Microcomb*, Nat Commun **10**, 680 (2019).
[32] C. Wang, M. Zhang, B. Stern, M. Lipson, and M. Lončar, *Nanophotonic Lithium Niobate Electro-Optic Modulators*, Opt Express **26**, 1547 (2018).
[33] C. Wang, M. Zhang, X. Chen, M. Bertrand, A. Shams-Ansari, S. Chandrasekhar, P. Winzer, and M. Lončar, *Integrated Lithium Niobate Electro-Optic Modulators Operating at CMOS-Compatible Voltages*, Nature **562**, 101 (2018).
[34] A. Herter, A. Shams-Ansari, F. F. Settembrini, H. K. Warner, J. Faist, M. Lončar, and I.-C. Benea-Chelmus, *Terahertz Waveform Synthesis in Integrated Thin-Film Lithium Niobate Platform*, Nat Commun **14**, 11 (2023).
[35] B. Chen, Z. Ruan, X. Fan, Z. Wang, J. Liu, C. Li, K. Chen, and L. Liu, *Low-Loss Fiber Grating Coupler on Thin Film Lithium Niobate Platform*, APL Photonics **7**, 076103 (2022).
[36] K. K. Mehta, C. D. Bruzewicz, R. McConnell, R. J. Ram, J. M. Sage, and J. Chiaverini, *Integrated Optical Addressing of an Ion Qubit*, Nat Nanotechnol **11**, 1066 (2016).
[37] K. K. Mehta and R. J. Ram, *Precise and Diffraction-Limited Waveguide-to-Free-Space Focusing Gratings*, Sci Rep **7**, 2019 (2017).
[38] E. Lomonte, F. Lenzini, and W. H. P. Pernice, *Efficient Self-Imaging Grating Couplers on a Lithium-Niobate-on-Insulator Platform at near-Visible and Telecom Wavelengths*, Opt Express **29**, 20205 (2021).
[39] C. Xiang, W. Jin, and J. E. Bowers, *Silicon Nitride Passive and Active Photonic Integrated Circuits: Trends and Prospects*, Photonics Res **10**, A82 (2022).
[40] D. Zhu et al., *Integrated Photonics on Thin-Film Lithium Niobate*, Adv Opt Photonics **13**, 242 (2021).
[41] N. Li, C. P. Ho, S. Zhu, Y. H. Fu, Y. Zhu, and L. Y. T. Lee, *Aluminium Nitride Integrated Photonics: A Review*, **10**, 2347 (2021).
[42] L. Splitthoff, M. A. Wolff, T. Grottke, and C. Schuck, *Tantalum Pentoxide Nanophotonic Circuits for Integrated Quantum Technology*, Opt Express **28**, 11921 (2020).
[43] D. Taillaert, P. Bienstman, and R. Baets, *Compact Efficient Broadband Grating Coupler for Silicon-on-Insulator Waveguides*, Opt Lett **29**, 2749 (2004).
[44] A. F. Oskooi, D. Roundy, M. Ibanescu, P. Bermel, J. D. Joannopoulos, and S. G. Johnson, *Meep: A Flexible Free-Software Package for Electromagnetic Simulations by the FDTD Method*, Comput Phys Commun **181**, 687 (2010).
[45] M. Maeda and H. Nakamura, *Hydrogen Bonding Configurations in Silicon Nitride Films Prepared by Plasma-enhanced Deposition*, J Appl Phys **58**, 484 (1985).
[46] S. Aachboun, P. Ranson, C. Hilbert, and M. Boufnichel, *Cryogenic Etching of Deep Narrow Trenches in Silicon*, Journal of Vacuum Science & Technology A **18**, 1848 (2000).
[47] M. J. Walker, *Comparison of Bosch and Cryogenic Processes for Patterning High-Aspect-Ratio Features in Silicon*, in *Proc.SPIE*, Vol. 4407 (2001), pp. 89–99.